\documentclass{article}
\newcommand {\apj}{Astrophys. J.}
\newcommand{\aap}{Astron. Astrophys.}
\newcommand{\prl}{Phys. Rev. Lett.}
\newcommand{\prd}{Phys. Rev. D}
\newcommand{\prc}{Phys. Rev. C}

\usepackage{epsfig,rotate,multicol,graphicx,psfrag,bm,times}
\begin{document}

\title{"Latent heat" of first-order varying pressure transitions  }
\author{Zheng Xiaoping\dag, Zhang Li\dag, Zhou Xia\dag, Kang Miao\ddag
\\
{\small \dag Department of Physics, Huazhong Normal University,
Wuhan430079 P. R. China}\\
{\small \ddag The college of physics and electron,Henan
university, Kaifeng475004, Henan, P.R.China }
}
\date{}
\maketitle
\begin{abstract}
We consider the energy release associated with first-order
transition by Gibbs construction and present such energy release
as an accumulation of a series of tiny binding energy differences
between over-compressed states and stable ones. Universal formulae
for the energy release from one homogeneous phase to the other is
given. We find the energy release per converted particle varies
with number density. As an example, the deconfinement phase
transition at supranuclear densities is discussed in detail. The
mean energy release per converted baryon is of order 0.1MeV in RMF
theory and MIT bag descriptions for hadronic matter and
 strange quark matter for a wider parameter region.

\end{abstract}
\section{Introduction}
Glendenning\cite{gle92} had  realized the essentially different
character of a first-order transition in simple system possessing
a single conserved quantity and complex one having more than one
conserved charge. One of the most remarkable features of a simple
system  is the constancy of the pressure during the transition
from one homogeneous phase to the other. Such phase transition is
the typical  description of first-order one in textbook. The
properties of the transition are quite different in complex. The
pressure varies continuously with the proportion of two pure
phases in equilibrium. Some quantities are obviously nonlinear
functions in the proportion in the mixed phase.

As well-known, the occurrence of first-order transition will
eventually accompany the release of energy. The energy is called
latent heat related to the release of entropy in the example of
gas-liquid transition. The phase transition at zero temperature
showed the liberation of binding energy. We still use the concept
of "latent heat" for the release of energy associated with such
the first-order phase transition.

The calculations of "latent heat" from hadronic matter to strange
quark matter were firstly made by Haensel and Zdunik\cite{hae91} .
They considered hadrons be absorbed into strange quark matter in
accreting strange stars and had expected that the heat per
absorbed neutron is in the range 10$\sim$30MeV. Generally the
energy release for constant-pressure phase transition at zero
temperature takes place nearby a transition point where an
over-compressed state, metastable, is probably the first to form
and then go to a stable state\cite{hae86}. However,  the phase
transition in a multicomponent system takes place in a varying
pressure region. An appearance of over-compressed state at every
pressure is possible. So we need to face up to a series of
metastable states  if the assumed system has a durative transfer
from one pure phase to the other. Relatively speaking, the
treatment of such energy release is a more difficult task than the
constant-pressure case. We will focus on solving the problem in
present paper.

Many phase transitions in nuclear matter are the expected ones
having multicomponent mixture in chemistry, the nuclear gas-liquid
transition relevant to accelerator experiments, where the neutron
and proton number are conserved, pion, koan condensates and
deconfinement of quarks in high density,  where the conserved
charges are baryon and electric. Typically, we will treat the
problem on the converted energy taking the confined-deconfined
phase transition as an example.

Ideally the two phase should be described selfconsistently in the
same theory, but we cannot solve QCD through confined-deconfined
phase transition based on our present knowledge. Hence we will use
separate models for each phase. We describe the equation of state
of hadronic matter in the framework of the relativistic mean-field
theory and depict the deconfined quark matter in the MIT bag
model. We will study the baryon number density dependence of
energy per baryon for different phases, hadronic, quark and their
mixed phase.

The plan of the paper is as follows. In Sec. II we state our views
on "latent heat" and formulate the problem. In Sec.III we solve
numerically the equations in the confined-deconfined phase
transition for the given equations of state. Our discussion and
conclusion will be made in Sec.IV.

\section{Formulation of the problem}
For superdense nuclear matter where a first-order phase transition
would take place, the pressure $p$ versus baryon number density
could generally be described in Figure 1 if we follow the same
philosophy and method used by Glendenning.
\begin{figure}
\centerline{\psfig{file=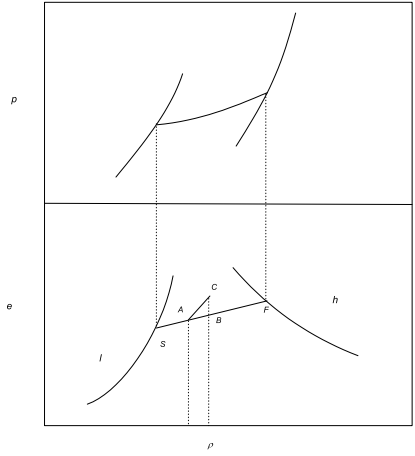,width=9cm}} \caption[]{The
schematic diagram for the equation of state having more than one
conserved charge}
\end{figure}
Correspondingly, the number density dependence of the energy per
baryon  could also be revealed in the figure(low panel). The body
experiences from low density phase, mixed phase to high density
phase with the number density increase. We find the suppression of
the increase binding energy  in the mixed phase as compared with the
low density phase when the body gradually compressed. The starting
point $S$ in the mixed phase is so conspicuous that the body may go
beyond $S$ in low density phase to an over-compressed state and then
have a transition to a stable state in mixed phase. Analogous
situation would appear at every point in mixed phase region. The
process depicted by $A, B$ and $C$ should repeatedly occur through
the duration when the body were transferred from $S$ to $F$. We can
consider that the transfer is nearly along the line $m$, denoting
the equation of state, if the over-compressed states have very small
deviations from the corresponding state in mixed phase region.

As well-known, the total energy  and baryon number densities for
the mixed phase, given by Glendenning\cite{gle92,gle97}, read
 \begin{equation}
\epsilon=\chi\epsilon_h+(1-\chi)\epsilon_l,
\end{equation}
\begin{equation}
\rho=\chi \rho_h+(1-\chi) \rho_l.
\end{equation}with the volume fraction
\begin{equation}
\chi=\frac{V_h}{V_l+V_h}
\end{equation}
where $V_l$ and $V_h$ represent the volumes occupied by low
density and high density phase respectively, $\epsilon_l$ and
$\epsilon_h$ denote  the corresponding energy densities, $\rho_l$
and $\rho_h$, the baryon number densities.

We further introduce the baryon number fraction $\eta (=N_h/N)$ to
rewrite the equations (1) and (2) for convenience of the following
treatments,where $N_h$ and $N$ present high density phase and
total baryon numbers.

Substitute $V_l=N_l/\rho_l, V_h=N_h/\rho_h$ into Eq(3), we get
\begin{equation}
\chi=\frac{\eta \rho_l}{\eta
\rho_l+(1-\eta)\rho_h}=\eta{\rho\over\rho_h}
\end{equation}
\begin{equation}
1-\chi=\frac{(1-\eta )\rho_h}{\eta
\rho_l+(1-\eta)\rho_h}=(1-\eta){\rho\over\rho_l}
\end{equation}

Therefore Eqs(1) and (2) become
\begin{equation}
\epsilon=\frac{\eta \rho_l}{\eta
\rho_l+(1-\eta)\rho_h}\epsilon_h+\frac{(1-\eta) \rho_h}{\eta
\rho_l+(1-\eta)\rho_h}\epsilon_l
\end{equation}
\begin{equation}
\rho=\frac{\rho_h\rho_l}{\eta \rho_l+(1-\eta)\rho_h}
\end{equation}

The energy per baryon(binding energy) is expressed as
\begin{equation}
e=\frac{\epsilon}{\rho}=\eta e_h+(1-\eta)e_l
\end{equation}
At zero temperature the energy $e$ only depends on the density.
The change in $e$ arises from the density increases by means of
compression and deconfinement.The energy $e$ can therefore be
assumed to be of the function of form $e(\eta(\rho),\rho)$.  We
find the derivative of $e$ with respect to $\rho$, which reads
\begin{equation}
{{\delta}e\over{\delta}\rho}=\left({{\partial}e\over{\partial}\rho}\right
)_\eta+{{\partial}e\over{\partial}\eta}{\delta\eta\over\delta\rho}.
\end{equation}
Furthermore
\begin{equation}
\delta e =e+{{\delta}e\over{\delta }\rho}
         = e+\left({{\partial}e\over{\partial}\rho}\right
)_\eta\delta\rho+{{\partial}e\over{\partial}\eta}\delta\eta,
\end{equation}
where
\begin{equation}
\left({{\partial}e\over{\partial}\rho}\right
)_\eta=\eta\left({{\partial}e_h\over{\partial}\rho}\right
)_\eta+(1-\eta )\left({{\partial}e_l\over{\partial}\rho}\right
)_\eta.
\end{equation}

Eq(10) shows the fact that the increase of binding energy in mixed
phase amounts to the composition of two processes, the simple
compressional change without phase transition and the
compressional one due to the occurrence of phase transition. We
define
\begin{equation}
-\delta q={\partial e\over\partial\eta}\delta\eta,
\end{equation}
and hence substitute it into Eq(9), we immediately have
\begin{equation}
\delta q =\left [\left(\frac{\partial e}{\partial\rho}\right
)_\eta-\frac{\delta e}{\delta\rho}\right ]\delta\rho=\left
[(e+\left({{\partial}e\over{\partial}\rho}\right )_\eta\delta\rho
)-( e+{{\delta}e\over{\delta }\rho}\delta\rho)\right]
\end{equation}

In equation(11), $\eta$ has been supposed invariable factor with
increase density. $\left(\frac{\partial e}{\partial\rho}\right
)_\eta$ indicates the the binding energy increase by a simple
compression regardless of phase transition since no particle is
converted to the other phase from one phase.  The second term of
the right-hand side in Eq(9) or Eq(12) is obviously associated
with phase transition because it arises from the change of the
baryon number fraction. We further change Eq(13) to reveal the
physical meanings of quantity $\delta q$. From an initial
state(see $A$ in figure 1) to the adjacent stable state, the
system suffering phase transition(see $B$ ), or the adjacent
metastable one, the system experiencing simple compression( see
$C$), through a equivalent density change, the binding energies
for $B$ and $C$ can be calculated as ${E_B\over
N}=e+{{\delta}e\over{\delta }\rho}\delta\rho$ and ${E_C\over
N}=e+\left ({{\delta}e\over{\delta }\rho}\right )_\eta\delta\rho$.
The $\delta q$ is immediately the energy difference between the
metastable state and the stable one, i.e.$\delta q={E_C\over
N}-{E_B\over N}$, where $E_B, E_C$ represent  total energies of
the system for the two states. Evidently $\delta q$ indicates the
binding energy release, just the same mechanism as the physical
situation in \cite{hae91}, when the body change from the
metastable state $C$ into the stable state$B$. A very small
 change of density considered here is the only difference from
 Ref\cite{hae91} and hence the energy release, $\delta q$, tends
 to zero. But the accumulation of a series of infinitesimal
 released energies is able to be nonvanishing because the system
 certainly experiences the finite change during real phase transition,
 which can be expressed by an definite integral.

Therefor, when we consider the assumed system with $N$ baryon
number from pure low density phase($S$ in figure 1) to pure high
density one($F$ in the figure), the  energy release can be
expressed as
\begin{equation}
Q=N\bar{q}=\int_{\rm S}^{\rm F} N\delta q =N\int_{\rm S}^{\rm F}
\left [\left (\frac{\partial e}{\partial\rho}\right
)_\eta-\frac{\delta e}{\delta\rho}\right ]\delta\rho,
\end{equation}
where $\bar{q}$ is mean energy release per baryon.  We can also
calculate the energy release per converted baryon for different
densities. The total energy release for given density $\rho$ in
mixed phase region equals $N\delta q$. Meanwhile, the converted
baryon numbers should be $\delta N_h$ at infinitesimal compression
$\delta\rho$. Thus the energy release per converted baryon is
calculated by
\begin{equation}
q(\rho)={N\delta q\over\delta N_{\rm h}}=\left ({{\delta
q\over\delta\rho}\over {{\rm d}\eta\over{\rm d} \rho}}\right )
\end{equation}
The meanings of the formula will be displayed thoroughly when
Eq(12) is substituted into Eq(15),
\begin{equation}
q(\rho)={\partial e\over\partial\eta}
\end{equation}
Clearly the energy release per converted baryon is different for
various baryon number density during phase transition in varying
pressure. We turn back to the understandings of Eq(10). We easily
find that the simple compression causes the binding energy
increase but the phase transition leads to the reduction. The
reduced energy is just about the energy release.
\section{Application: The energy release in deconfinement phase transition }

We will numerically solve the integral(12) together with
equations(1)and(2)  for the transition from hadronic matter to
quark matter. The low density phase represents hadronic matter,
called HP for short,  and the high density one indicates quark
matter, QP for short. We  use the equations of state in
relativistic mean-field theory(RMF) description for hadronic
matter and the equations of state in MIT bag model for quark
matter.
\begin{eqnarray}
 \textit{ \L}=\sum_{B}\overline{\psi}_{B}(i\gamma_{\mu}\partial^{\mu}-m_{B}+g_{\sigma B}\sigma-g_{\omega B}\gamma_{\mu}\omega^
{\mu}-\frac{1}{2}g_{\rho B}\gamma_{\mu}\tau\cdot\rho^{\mu})\psi_{B}\nonumber \\
+\frac{1}{2}(\partial_{\mu}\sigma\partial^{\mu}\sigma-m_{\sigma}^{2}\sigma^{2})+
\sum_{\lambda=e,\mu}\overline{\psi}_{\lambda}(i\gamma_{\mu}\partial^{\mu}-m_{\lambda})\psi_{\lambda}\nonumber \\
-\frac{1}{4}\omega_{\mu\nu}\omega^{\mu\nu}+\frac{1}{2}m_{\omega}^{2}\omega_{\mu}\omega^{\mu}-
\frac{1}{4}\rho_{\mu\nu}\cdot\rho^{\mu\nu}\nonumber \\
+\frac{1}{2}m_{\rho}^{2}\rho_{\mu}\cdot\rho^{\mu}-\frac{1}{3}bm_{n}(g_{\sigma
n}\sigma)^{3}-\frac{1}{4}c(g_{\sigma n}\sigma)^{4}.
\end{eqnarray}
where the spinor for the baryon species B is denoted by
$\psi_{B}$, $\sigma, \omega, \rho$ represent meson fields. The
energy density and pressure can be obtained in the familiar manner
as
\begin{eqnarray}\label{n07}
  \epsilon_{HP}=\frac{1}{3}bm_{n}(g_{\sigma}\sigma)^{3}+\frac{1}{4}c(g_{\sigma}\sigma)^{4}
  +\frac{1}{2}m_{\sigma}^{2}\sigma^{2}+\frac{1}{2}m_{\omega}^{2}\omega_{0}^{2}
  +\frac{1}{2}m_{\rho}^{2}\rho_{03}^{2}\nonumber \\
  +\sum_{B}\frac{2J_{B}+1}{2\pi^{2}}\int_{0}^{k_{B}}\sqrt{k^{2}+(m_{B}-g_{\sigma B}\sigma)^{2}}
  k^{2}dk \nonumber \\
  +\sum_{\lambda}\frac{1}{\pi^{2}}\int_{0}^{k_{\lambda}}\sqrt{k^{2}+m_{\lambda}^{2}}k^{2}dk  .
\end{eqnarray}
\begin{eqnarray}\label{n08}
 p_{HP}=-\frac{1}{3}bm_{n}(g_{\sigma}\sigma)^{3}-\frac{1}{4}c(g_{\sigma}\sigma)^{4}
  -\frac{1}{2}m_{\sigma}^{2}\sigma^{2}+\frac{1}{2}m_{\omega}^{2}\omega_{0}^{2}
  +\frac{1}{2}m_{\rho}^{2}\rho_{03}^{2}\nonumber \\
  +\frac{1}{3}\sum_{B}\frac{2J_{B}+1}{2\pi^{2}}\int_{0}^{k_{B}}\frac{k^{4}}{\sqrt{k^{2}+(m_{B}-g_{\sigma B}\sigma)^{2}}}
 dk \nonumber \\
 +\frac{1}{3}\sum_{\lambda}\frac{1}{\pi^{2}}\int_{0}^{k_{\lambda}}\frac{k^{4}}{\sqrt{k^{2}+m_{\lambda}^{2}}}dk .
\end{eqnarray}
where the coupling constants
$g_{\sigma}$,$g_{\omega}$,$g_{\rho}$,$b$ and $c$ can be determined
by the nuclear saturation density $\rho_{0}$, the binding energy
at saturation $\textit{B/A}$, the symmetry energy$a_{sym}$, the
compression modulus$\textit{K}$ and the effective nucleon mass
$m^{*}$. In MIT bag model, the energy density and pressure of
quark matter at zero temperature are given by
\begin{equation}\label{n04}
  \epsilon_{QP}=B+\sum_f\frac{3}{4\pi^{2}}\left [\mu_f(\mu_f
^{2}-m_f^{2})^{1/2}(\mu_f^2-{1\over 2}m_f^2)-{1\over 2}m_f^{4}\ln
{\mu_f+(\mu_f^2-m_f^2)^{1/2} \overwithdelims() m_f} \right ].
\end{equation}
\begin{equation}\label{n05}
p_{QP}=-B+\sum_f\frac{1}{4\pi ^{2}}\left [\mu_f(\mu_f
^{2}-m_f^{2})^{1/2}(\mu_f^2-{5\over 2}m_f^2)+{3\over 2}m_f^{4}\ln
{\mu_f+(\mu_f^2-m_f^2)^{1/2} \overwithdelims() m_f} \right ].
\end{equation}

We calculate the mixed phase of hadronic and quark matter with
Gibbs construction which first used by Glendenning\cite{gle92}.
For simplicity, we neglect Coulom and surface effects in MP. The
Gibbs condition for mechanical and chemical equilibrium at zero
temperature in MP is written as
\begin{equation}
p_{\rm PH}(\mu_n,\mu_e)=p_{\rm QP}(\mu_n,\mu_e)
\end{equation}
We will obtain Eqs(1) and (2) by applying the condition (20)
together with Eqs(16)$\sim$ (19) and imposing the condition of
global charge neutrality in MP

\begin{equation}
\chi\rho^c_{\rm QP}+(1-\chi)\rho^c_{\rm HP}=0,
\end{equation}
where $\rho^c_{\rm QP}$ and $\rho^c_{\rm HP}$ respectively denote
negative charged density of quark matter and positive one of
hadronic matter. When $\chi$=0 and 1, the global charge neutrality
spontaneously come back local charge neutralities for pure
hadronic matter and pure quark matter. The equilibrium chemical
potentials can be determined in MP from Eq(20) while $\chi$ can be
solved from Eq(21).

We choose the representative parameters for soft, moderate and
stiff hadronic matter equations of state listed in table 1. The
bag constant is taken as $B^{1/4}$=170MeV, 180MeV and 190MeV. We
construct a series of the equations of state with MP denoted by
RMFn+$B^{1/4}$(n=1,2,3)  under those considerations. As an
example, RMF1+180 is depicted in figure 2.
\tabcolsep 4pt
\begin{table}[4pt,b]
\caption[]{Nucleon-meson coupling constants}
\begin{center}\begin{tabular}{ccccccc} \hline
 Name  & $({g_{\sigma}\over m_{\sigma}})^{2}({\rm fm}^{2})$ &  $({g_{\omega}\over m_{\omega}})^{2}({\rm fm}^{2})$ &
$({g_{\rho}\over m_{\rho}})^{2}({\rm fm}^{2})$ & $100b$ & $100c$ & Ref\\
\hline
 RMF1  & 11.79 & 7.149 & 4.411 & 0.2947 & -0.1070 & \cite{gle97}\\
 RMF2  & 8.492 & 4.356 & 5.025 & 0.2084 & 2.780 & \cite{gho95}\\
 RMF3  & 10.339 & 4.820 & 4.791 & 1.1078 & -0.9751 & \cite{gle97}\\ \hline
\end{tabular}\end{center}
\end{table}

We calculates ${\delta q\over\delta\rho}$ and $\bar{q}$ for the
different equations of state. The results are showed in figure 3
and 4 . Although the uncertainties of the equations of state have
effect on the results, they are of order
 0.1MeV under our considerations.

 We also utilize equation (15) to calculate $q({\rho})$ shown in figure 5 and 6,
corresponding to figure 3 and 4. Obviously, $q({\rho})$ changes
around $\bar{q}$ in the wide regions. The data for several cases,
RMF1+180, RMF2+180 and RFM3+180, are listed in tables 2$\sim$4.
\begin{figure}
\centerline{\psfig{file=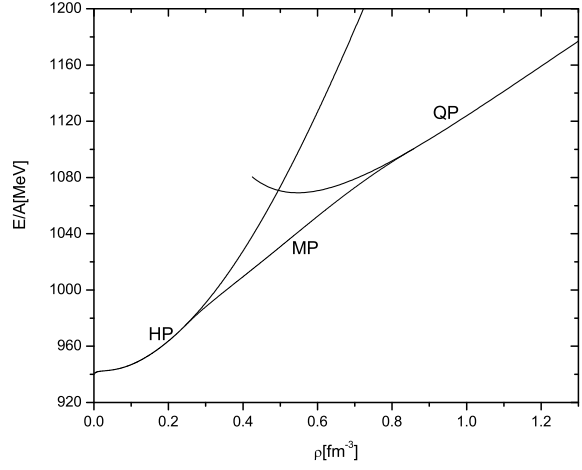,width=9cm}} \caption[]{Energy per
baryon versus baryon number density for RMF1+180}
\end{figure}
\begin{figure}
\centerline{\psfig{file=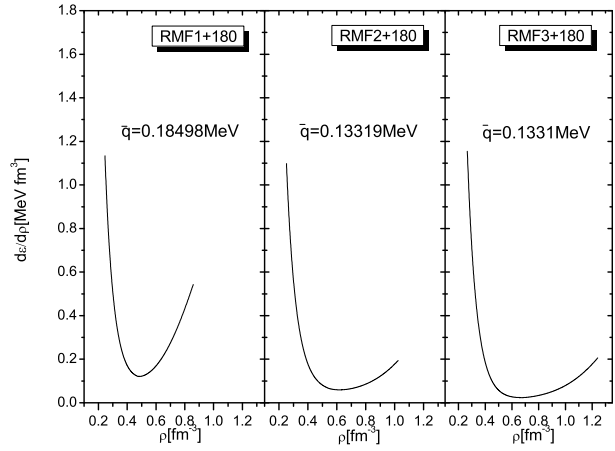,width=9cm}} \caption[]{The
${\delta q\over\delta\rho}$ versus $\rho$ for soft, moderate and
stiff hadronic matter equation of state. The bag constant is given
$B^{1/4}=180$MeV}
\end{figure}
\begin{figure}
\centerline{\psfig{file=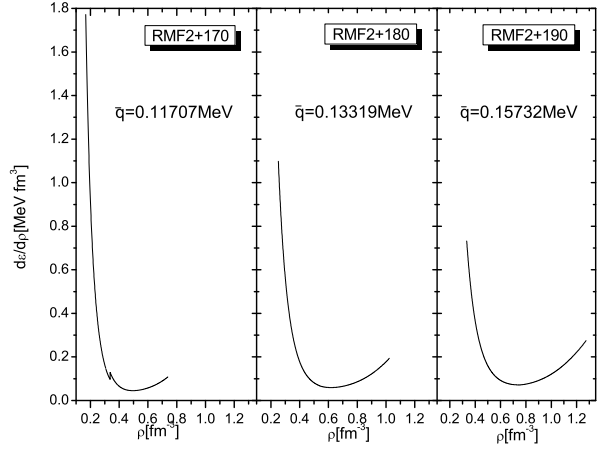,width=9cm}} \caption[]{The
${\delta q\over\delta\rho}$ versus $\rho$ for different bag
constant models. The hadronic matter equation of state is given as
moderate soft one.}
\end{figure}
\begin{figure}
\centerline{\psfig{file=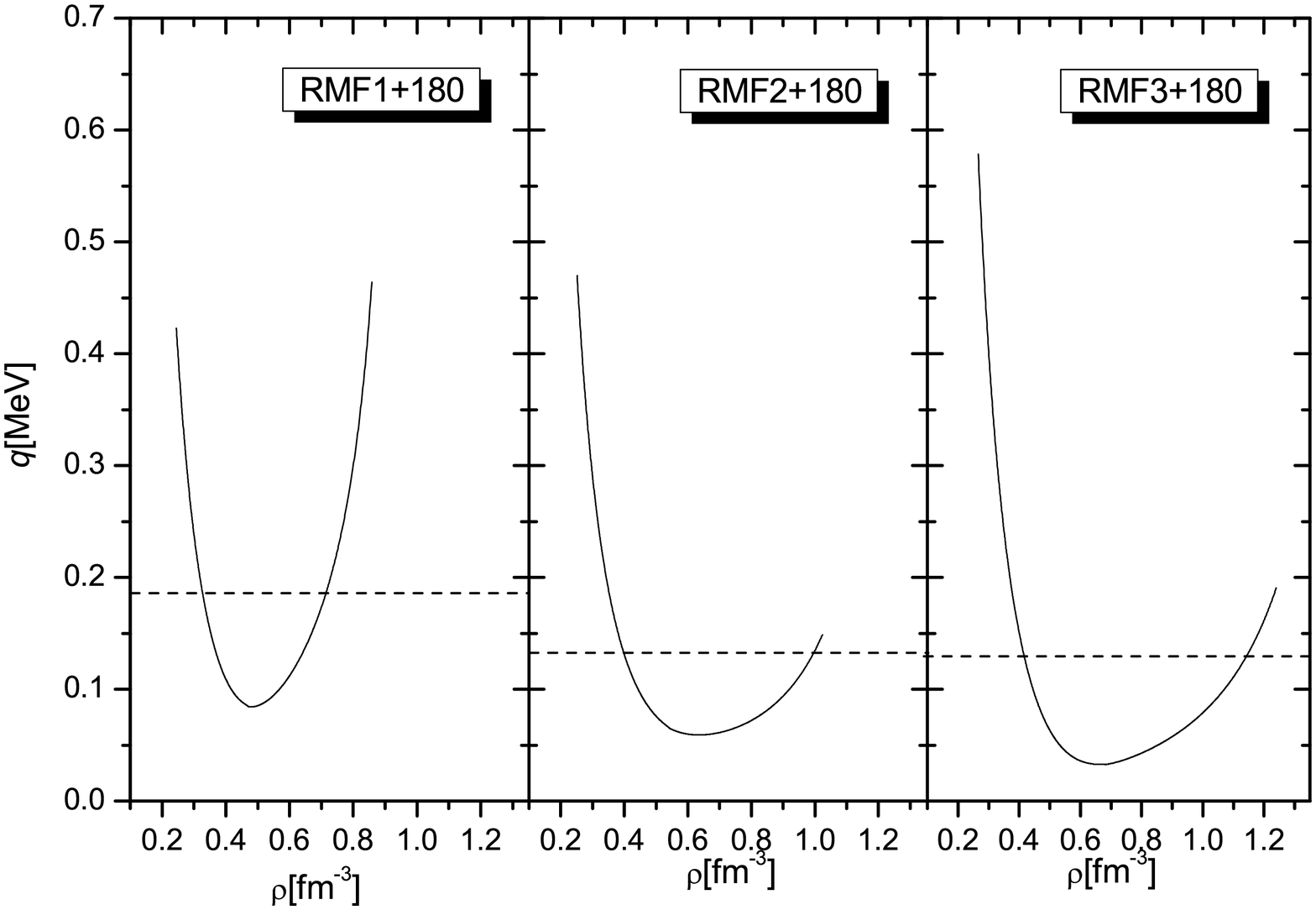,width=9cm}} \caption[]{The baryon
number density dependence of releasing energy per a converted
baryon  for soft, moderate and stiff hadronic matter equation of
state. The horizontal lines represent the mean values.}
\end{figure}

\begin{figure}
\includegraphics[width=0.7\textwidth,bb=0 0 818 572] {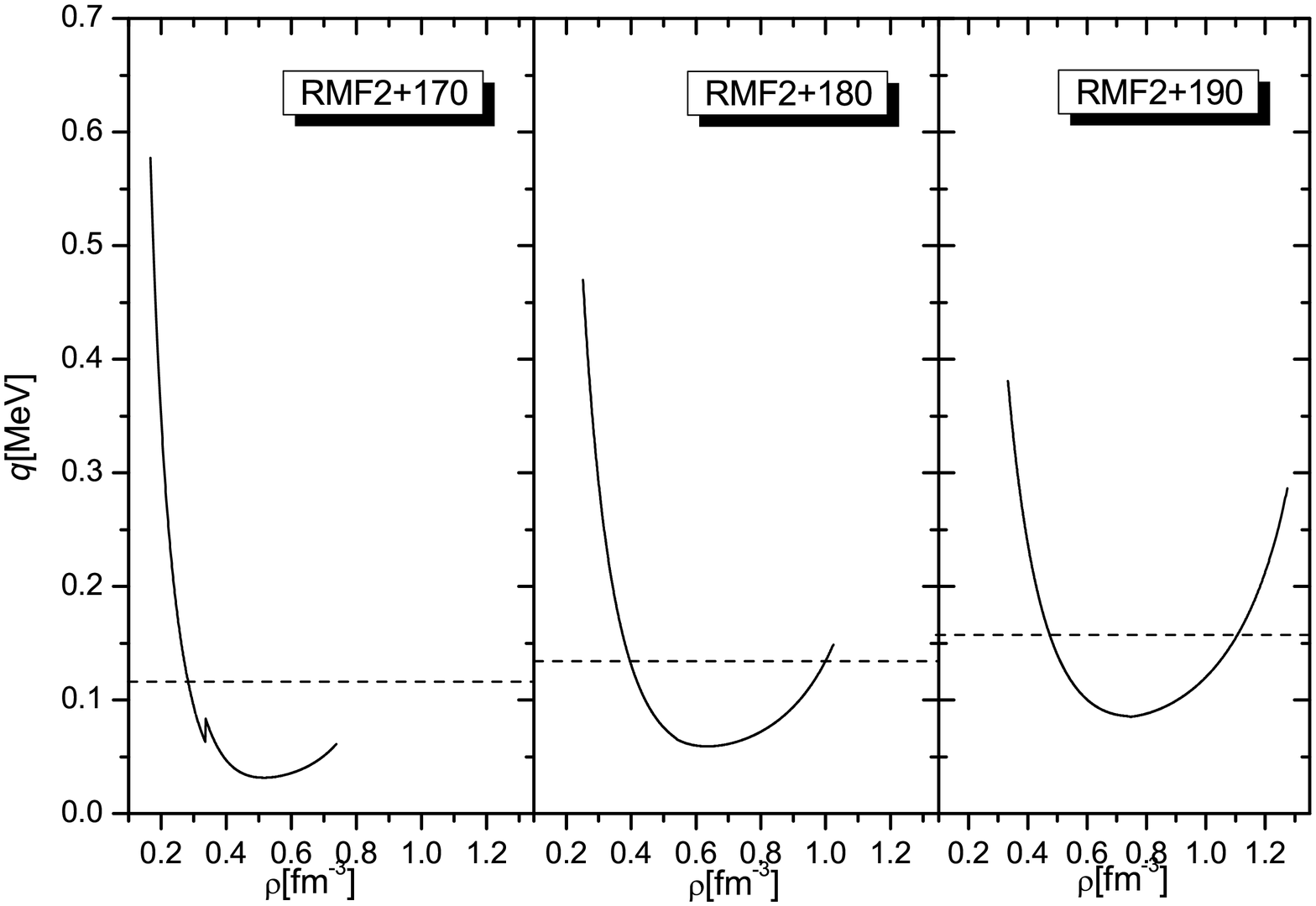}
\caption[]{The baryon number density dependence of releasing
energy per a converted baryon for different bag constants. The
horizontal lines represent the mean values.}
\end{figure}

\section{Conclusion and discussion}

We put forward a method to calculate the energy release in the
first-order phase transition that have varying pressure found by
Glendenning. We formulate the release of energy per baryon from
one homogenous phase to the other. We find the dependence of
energy release per converted baryon on density.

We investigate the case of deconfinement phase transition using
RMF description for hadronic matter and MIT bag description for
quark matter. We find that the mean energy release per converted
baryon is of order 0.1MeV, much smaller than that of the
transition considered by Haensel and Zdunik, tens of
MeV\cite{hae91}. We indeed see the difference of energy release if
a hadron is converted into quarks at different pressure(density)
points.

From the analysis and the performance of mathematical
calculations, we see that the binding energy in a hadron is
probably liberated during transition process at zero temperature.
Clearly the degree of freedom for a baryon number changes in
deconfinement phase transition. Our calculations also show that
the deconfinement process in varying pressure can be regard as an
accumulation of innumerable constant-pressure phase transitions.

Our discussion may be relevant to accelerator experiments and many
astrophysical problems. Especially, many transitions with two
conserved charges could occur in neutron stars, such as nuclei
compositional transition in the crust\cite{iid97}, meson
condensates\cite{gle99}, hyperon productions and superfluid
transitions in the interior\cite{gia04}. The consideration in this
work may be applied to them. The energy release due to such
first-order transitions would significantly influence the
evolution of neutron stars\cite{iid97,kan07, kan08}.

This work is supported by NFSC under Grant Nos.10773004, 10603002
 and 10747126.

\begin{table}[4pt,b]
\caption[]{The calculations of $q({\rho})$corresponding to the
equation of state of RMF1+180}
\begin{center}\begin{tabular}{ccccccc} \hline
 $\rho (fm^{-3})$ &  $P(MeVfm^{-3})$ & $\eta$ & $q(\rho)(MeV)$ \\
\hline
0.24562 & 15.99590 & 0.00252 & 0.419183681 \\
0.28521 & 18.92140 & 0.10374 & 0.275781000 \\
0.32505 & 22.89477 & 0.18689 & 0.189162009 \\
0.36542 & 28.14901 & 0.25904 & 0.136300761 \\
0.40539 & 34.62885 & 0.32302 & 0.105876989 \\
0.42517 & 38.28373 & 0.35297 & 0.096513344 \\
0.44539 & 42.28858 & 0.38282 & 0.089926622 \\
0.46579 & 46.57121 & 0.41241 & 0.085636798 \\
0.48530 & 50.85101 & 0.44036 & 0.084306666 \\
0.49537 & 53.10652 & 0.45483 & 0.084855744 \\
0.50556 & 55.40860 & 0.46950 & 0.085843691 \\
0.51543 & 57.65472 & 0.48379 & 0.087173365 \\
0.52501 & 59.83958 & 0.49772 & 0.088797472 \\
0.53523 & 62.17201 & 0.51266 & 0.090895052 \\
0.54519 & 64.44081 & 0.52732 & 0.093310771 \\
0.55540 & 66.75197 & 0.54243 & 0.096084463 \\
0.56539 & 68.99347 & 0.55730 & 0.099108422 \\
0.57515 & 71.16039 & 0.57192 & 0.102399167 \\
0.58520 & 73.36496 & 0.58709 & 0.106045033 \\
0.59505 & 75.48930 & 0.60201 & 0.109986619 \\
0.60524 & 77.64971 & 0.61755 & 0.114346182 \\
0.61583 & 79.84750 & 0.63378 & 0.119247950 \\
0.63516 & 83.72478 & 0.66360 & 0.129212059 \\
0.65549 & 87.59585 & 0.69515 & 0.141124493 \\
0.67551 & 91.18872 & 0.72634 & 0.154501478 \\
0.69505 & 94.47733 & 0.75679 & 0.169333315 \\
0.71571 & 97.72125 & 0.78892 & 0.187330950 \\
0.73563 & 100.6232 & 0.8197  & 0.207391595 \\
0.75550 & 103.3067 & 0.85015 & 0.230686953 \\
0.77510 & 105.7551 & 0.87984 & 0.257760827 \\
0.79554 & 108.1105 & 0.91037 & 0.291723249 \\
0.81526 & 110.2041 & 0.93934 & 0.332089997 \\
0.83551 & 112.1845 & 0.96855 & 0.384646739 \\
0.85620 & 114.0450 & 0.99778 & 0.447912000 \\\hline
\end{tabular}\end{center}
\end{table}

\begin{table}[4pt,b]
\caption[]{The calculations of $q({\rho})$corresponding to the
equation of state of RMF2+180}
\begin{center}\begin{tabular}{ccccccc} \hline
 $\rho (fm^{-3})$ &  $P(MeVfm^{-3})$ & $\eta$ & $q(\rho)(MeV)$ \\
\hline
0.25134 & 15.73143 & 0.00151 & 0.469751750 \\
0.29175 & 18.45115 & 0.09636 & 0.313446165 \\
0.33167 & 21.94685 & 0.17143 & 0.219698869 \\
0.37161 & 26.38217 & 0.23424 & 0.160179216 \\
0.41056 & 31.70503 & 0.28737 & 0.122075922 \\
0.44546 & 37.35738 & 0.33025 & 0.098760648 \\
0.45562 & 39.15959 & 0.34209 & 0.093479673 \\
0.46561 & 41.00209 & 0.35350 & 0.088757745 \\
0.47594 & 42.97941 & 0.36510 & 0.084349887 \\
0.48566 & 44.90266 & 0.37581 & 0.080640911 \\
0.49573 & 46.96361 & 0.38676 & 0.077251319 \\
0.50570 & 49.06666 & 0.39744 & 0.074219457 \\
0.51558 & 51.21138 & 0.40790 & 0.071506976 \\
0.52583 & 53.50250 & 0.41863 & 0.069013661 \\
0.53555 & 55.73141 & 0.42870 & 0.066841209 \\
0.54566 & 58.11011 & 0.43906 & 0.064591642 \\
0.55527 & 60.42206 & 0.44884 & 0.063371717 \\
0.56576 & 63.00044 & 0.45949 & 0.062230485 \\
0.57578 & 65.51184 & 0.46965 & 0.061360053 \\
0.58580 & 68.06732 & 0.47980 & 0.060609410 \\
0.59582 & 70.66685 & 0.48995 & 0.060077471 \\
0.60540 & 73.18941 & 0.49966 & 0.059635930 \\
0.61546 & 75.87528 & 0.50989 & 0.059324310 \\
0.62555 & 78.60550 & 0.52018 & 0.059157773 \\
0.63570 & 81.38022 & 0.53056 & 0.059148265 \\
0.64544 & 84.07060 & 0.54056 & 0.059176934 \\
0.65524 & 86.80200 & 0.55068 & 0.059341975 \\
0.67079 & 91.17776 & 0.56685 & 0.059817394 \\
0.69090 & 96.89657 & 0.58802 & 0.060746128 \\
0.71096 & 102.6439 & 0.60947 & 0.062017483 \\
0.73053 & 108.2654 & 0.63074 & 0.063615371 \\
0.75066 & 114.0416 & 0.65302 & 0.065581225 \\
0.77091 & 119.8231 & 0.67590 & 0.067979420 \\
0.81089 & 131.0503 & 0.72245 & 0.073798485 \\
0.85070 & 141.8468 & 0.77076 & 0.081456043 \\
0.89050 & 152.1159 & 0.82099 & 0.091154274 \\
0.93036 & 161.7624 & 0.87312 & 0.103677438 \\
0.97023 & 170.6913 & 0.92685 & 0.119589566 \\
1.02250 & 181.2445 & 0.99918 & 0.148039000 \\

\hline
\end{tabular}\end{center}
\end{table}

\begin{table}[4pt,b]
\caption[]{The calculations of $q({\rho})$corresponding to the
equation of state of RMF3+180}
\begin{center}\begin{tabular}{ccccccc} \hline
 $\rho (fm^{-3})$ &  $P(MeVfm^{-3})$ & $\eta$ & $q(\rho)(MeV)$ \\
\hline
0.26621 & 15.38213 & 0.00033 & 0.578528090 \\
0.30013 & 17.25338 & 0.07188 & 0.401621596 \\
0.35009 & 20.96879 & 0.15573 & 0.242218167 \\
0.39063 & 25.04845 & 0.21094 & 0.164227120 \\
0.41041 & 27.45226 & 0.23484 & 0.136629769 \\
0.43008 & 30.14255 & 0.25704 & 0.114235389 \\
0.44629 & 32.59680 & 0.27429 & 0.098882871 \\
0.46224 & 35.23192 & 0.29047 & 0.086101857 \\
0.47848 & 38.14854 & 0.30624 & 0.075092011 \\
0.49401 & 41.16360 & 0.32073 & 0.066246291 \\
0.50990 & 44.47685 & 0.33500 & 0.058676162 \\
0.52612 & 48.10325 & 0.34909 & 0.052223703 \\
0.54224 & 51.94964 & 0.36267 & 0.047056820 \\
0.56641 & 58.15899 & 0.38235 & 0.041151896 \\
0.58231 & 62.52432 & 0.39493 & 0.038316741 \\
0.60639 & 69.53827 & 0.41355 & 0.035288569 \\
0.62200 & 74.32911 & 0.42539 & 0.034077900 \\
0.64609 & 82.06479 & 0.44342 & 0.033020453 \\
0.66213 & 87.42851 & 0.45532 & 0.032808586 \\
0.68627 & 95.78867 & 0.47311 & 0.032873272 \\
0.70224 & 101.4816 & 0.48487 & 0.033734009 \\
0.72612 & 110.1944 & 0.50259 & 0.035371294 \\
0.74231 & 116.2149 & 0.51474 & 0.036785477 \\
0.76631 & 125.2663 & 0.53299 & 0.039089761 \\
0.78216 & 131.3051 & 0.54525 & 0.040836874 \\
0.80619 & 140.5170 & 0.56417 & 0.043644712 \\
0.82229 & 146.7050 & 0.57712 & 0.045721299 \\
0.86230 & 162.0314 & 0.61029 & 0.051421958 \\
0.90207 & 177.0247 & 0.64485 & 0.058017367 \\
0.94236 & 191.7758 & 0.68158 & 0.065666026 \\
0.98239 & 205.8220 & 0.71984 & 0.074607477 \\
1.02218 & 219.0478 & 0.75958 & 0.084907678 \\
1.06237 & 231.5507 & 0.80132 & 0.097255920 \\
1.10218 & 243.0187 & 0.84405 & 0.111737914 \\
1.14205 & 253.5626 & 0.88802 & 0.129273054 \\
1.18264 & 263.3225 & 0.93368 & 0.150868505 \\
1.23835 & 275.1766 & 0.99721 & 0.189418000 \\
\hline
\end{tabular}\end{center}
\end{table}

\end{document}